# Broadband super-resolving lens with high transparency for propagating and evanescent waves in the visible range

Mark J. Bloemer[a)], Giuseppe D'Aguanno, Nadia Mattiucci, and Michael Scalora
*Dept. of the Army, Charles M. Bowden Facility, Bldg. 7804*
*Research, Development, and Engineering Command, Redstone Arsenal, AL 35898*

Neset Akozbek
*Time Domain Corp., Cummings Research Park, 7057 Old Madison Pike, Huntsville, AL 35806*

[a)]Electronic mail: mark.bloemer@us.army.mil

We present a theoretical analysis of a super-resolving lens based on 1-dimensional metallo-dielectric photonic crystals composed of Ag/GaP multilayers. The lens contains a total of 10 optical skin depths of Ag, yet maintains a normal incidence transmittance of ~50% for propagating waves over the super-resolving wavelength range of 500-650 nm. The individual Ag layers are 22 nm thick and can be readily fabricated in conventional deposition systems. The importance of anti-reflection coatings for the transmission of evanescent and propagating waves is illustrated by comparison to periodic and symmetric structures without the coatings. In addition, the reflection for propagating waves is reduced to ~5% across the super-resolving wavelength band diminishing the interference between the object and the lens.

In the year 2000, Pendry[1] showed that a simple metal film displayed negative refraction of the Poynting vector for TM polarized light and was capable of supporting evanescent waves. This unique combination resulted in a flat superlens that could image an object with a resolution beyond traditional glass lenses which support only propagating waves. The performance limit of the metallic superlens was associated with the losses in the metallic film.

In order to overcome the losses associated with a single metal film, Ramakrishna *et al.* designed a superlens based on a multilayer metal-dielectric stack having thin metal layers[2]. This new combination of a positive and negative dielectric constant material results in a slightly different type of superlens. Instead of having a focus inside of the metal film, the new geometry balanced the negative and positive refraction of the Poynting vector in consecutive layers. This leads to a waveguide-type effect that carries the propagating and evanescent waves through the structure without the usual diffraction.

Since the introduction of the metallo-dielectric superlens there have been numerous attempts to design structures with high transmittance for the evanescent waves[3-6]. However, very little has been said about the transmittance of the propagating waves and the bandwidth of the lens. In the following we take a slightly different approach to the problem and start with a metallo-dielectric structure known to have broadband and high transmittance for the propagating waves and ask, "What are the transmission characteristics for the evanescent waves?"

The structure we examine is a 1-dimensional metallo-dielectric photonic crystal (1-D MDPC) based on Ag/GaP. These 1-D MDPC are known to have a very low sheet resistance (~0.1 ohm/sq) due to the high fraction of metal in the multilayer stack[7]. In general, most of the interest in photonic crystals has concentrated on the unique properties of the band gap and the band edges, but for MDPC the interest is on the pass band precisely because of the unusual combination of broadband, high transmittance and high electrical conductivity, hence they are often called "transparent metals."

The concept of a transparent MDPC is based on a series of strongly coupled metal-dielectric Fabry-Perot cavities[8]. As an example, we start with a single cavity of Ag/GaP/Ag (22 nm/35 nm/22 nm) surrounded by air. Normally, the lowest order transmission resonance occurs at the half-wave condition, however, the phase change upon reflection needs to be explicitly included in the interference equation[9]. The departure from a perfect reflector can be substantial and in the case of our Ag/GaP/Ag cavity, the optical thickness of the GaP is closer to a quarter-wave. The normal incidence transmittance for the Fabry-Perot cavity shown in Fig. 1 displays a resonance at a wavelength of 570 nm. The transmittance was calculated using a standard matrix transfer method and the optical constants of Ag and GaP were taken from Palik[10].

Adding another period of GaP/Ag results in 2 coupled cavities and removes the degeneracy causing the single resonance at 570 to split into 2 resonances at 490 nm and 620 nm. The amount of the splitting depends on the thickness of the Ag layers, with thinner layers providing a larger separation in the transmission resonances. Adding another period results in 3 coupled cavities and 3 resonances and so on for more periods. Fig. 1 shows the transmittance for 5 coupled cavities or 5.5 periods of Ag/GaP (22 nm/35 nm). The details of the transmittance are strongly influenced by the dispersion of the Ag[7]. The most important feature to note in comparing the transmittance of the single cavity and 5 coupled cavities is the high degree of transparency in spite of the 5 cavity structure having ~10 optical skin depths of metal compared to ~3.5 optical skin depths for the single cavity. The high transparency is attributed to resonance tunneling through coupled cavities. In Ref. 11, a 1-D MDPC design was shown to have 20% transmittance across a 100 nm wide band in the visible region with a total of ~40 optical skin depths of Ag!

While the transmittance of the 5.5 period MDPC is large, it can be improved by the addition of antireflection coatings. It was found that the antireflection (AR) coating designed for a single metal-dielectric Fabry-Perot cavity[12] worked equally well for coupled cavities[11]. The AR coating is simply ½ the thickness of the dielectric spacer and of the same material. The transmittance of the 5.5 period MDPC with 17 nm thick GaP AR coatings is shown in Fig. 1. The AR coatings double the overall transmittance and smooth out the oscillations associated with the coupled cavity modes. Though not shown in Fig. 1, the reflectance in the pass band for the sample without the AR coatings is oscillatory from 65% to 15% with the low reflectance at the transmittance maxima and high reflectance at the transmittance minima. In contrast, the AR coatings reduce the oscillations in the reflectance to high values of only 10% and low values of 1%. For a more complete comparison, we plot the transmittance of a periodic structure in Fig. 1 containing 6 periods of Ag/Gap (22 nm/35 nm).

The angular dependence of the transmittance spectra in Fig. 1 depend strongly on the refractive index of the dielectric spacer. As the angle of incidence increases, the pass band shifts towards the blue, but the large refractive index of GaP (~3.5) reduces the change in the optical path with angle by causing the rays to stay close to the normal. The high refractive index also improves the overall transmittance of the MDPC. There is a complicated design procedure to optimize the transmittance through a single metal film[13]. For optimizing the transmittance of multiple cavity metal-dielectric filters we quote from p. 326 of MacCleod's book on thin-film optical filters[13], *"The accurate design procedure for such metal-dielectric filters can be* *lengthy and tedious and frequently they are simply designed by trial and error as they are manufactured."*

In the following we will examine the transmission characteristics for the evanescent waves and the super-lensing properties of MDPC. We compare 3 geometries, a Periodic MDPC [6 periods of Ag/Gap (22 nm/35 nm)], a Symmetric MDPC [5.5 periods of Ag/GaP (22 nm/35 nm)], and a Transparent Metal MDPC [5.5 periods of Ag/Gap (22 nm/35 nm) with 17 nm thick GaP AR coatings on the entrance and exit faces]. All three structures have the same amount of Ag, 132 nm total thickness equally divided into 6 layers. Similar to the transmittance spectra of Fig. 1 for the propagating waves, the 3 different MDPC have drastically different transmittance for the evanescent waves due to the first and last layers of the structures.

The calculations of the lensing properties of MDPC have been carried out using the technique of the angular spectrum decomposition [14] in conjunction with the transfer matrix technique. A plane, monochromatic, TM polarized wave with wave-vector $k_o$ in free space is incident on the object plane which is at a distance $d$ from the input surface of a MDPC lens of length $L$. Note that both the object plane and the lens are in free space. The magnetic field is expressed as:

$$\vec{H}(x,z,t) = (1/2)\left[\vec{\tilde{H}}(x,z)e^{-i\omega t} + c.c\right],$$

where $\vec{\tilde{H}}(x,z) = \tilde{H}(x,z)\hat{y}$ is the complex, stationary vector field, $\hat{y}$ is the unit vector of the y-axis and the z-axis is normal to the layers of the MDPC. The complex amplitude of the magnetic field $\tilde{H}(x,z)$ at $z \geq L$ (i.e. in the semi-space at the output of the lens) is expressed by the following integral:

$$\tilde{H}(x, z \geq L) = \int_{-\infty}^{+\infty} A(k_x) t(k_x) e^{ik_x x} e^{i\sqrt{k_0^2 - k_x^2}(z+d-L)} dk_x.$$

Here $A(k_x)$ is the Fourier spectrum of the magnetic field on the object plane, and $t(k_x)$ is the complex transmission function of the layered lens for TM polarization. The transmission function has been calculated using a matrix transfer technique [15].

The diffraction limited field is obtained by removing the evanescent components and extending the integral only over the propagating modes, i.e.

$$\tilde{H}_{dif\,\lim}(x, z \geq L) = \int_{-k_0}^{+k_0} A(k_x) t(k_x) e^{ik_x x} e^{i\sqrt{k_0^2 - k_x^2}(z+d-L)} dk_x.$$

The electric field at z≥L has been calculated from the magnetic filed using the relation: $\nabla \times \vec{\tilde{H}} = -i\omega\vec{\tilde{E}}$. Finally, the time averaged Poynting vector is obtained through the following formula, $\vec{S} = (1/2)\text{Re}[\vec{\tilde{E}}^* \times \vec{\tilde{H}}]$.

The transmittance of the MDPC is plotted in Fig. 2a as a function of $k_x/k_o$ at a wavelength of 532 nm and in Fig. 2b as a function of the wavelength for a value of $k_x/k_o$=3. Of course the ideal lens would have a transmittance of 100% for all wavelengths and all values of $k_x/k_o$. In reality, a lens is tailored for the specific application, meaning the lens should have good transmittance properties over the range of wavelengths and Fourier components consistent with the input spectrum. Looking at Fig. 2a we see that the symmetric MDPC has a very high transmittance for Fourier component just beyond $k_x/k_o$ =1, but it occurs over a very limited range. The periodic MDPC has a slightly broader transmittance for the evanescent components with a peak at $k_x/k_o$=1.25. In comparison, the transparent metal MDPC has a much broader range of $k_x/k_o$ values with good transmittance.

Fig. 2b gives an indication of the broadband nature of the transparent metal MDPC lens. Plotted is the transmittance for $k_x/k_o$=3. Without question, the transparent metal MDPC has a significantly higher transmittance for this particular Fourier component compared with the symmetric or periodic MDPC.

Fig. 3 is a 3-D plot of the transmittance versus wavelength and $k_x/k_o$. To save space we plot the 3-D transmittance for the transparent metal MDPC and the periodic MDPC. The symmetric MDPC is only slightly better than the periodic MDPC. The notable feature in Fig. 3 is the broadband, high transmittance of the propagating and evanescent waves for the transparent metal MDPC. The five resonances associated with the five coupled cavities are clearly evident. It is surprising that the AR coatings can have such a drastic effect on the transmittance of the propagating and evanescent waves over a broad range of wavelengths.

The super-lensing properties were examined by imaging 2 slits separated by less than $\lambda/2$ in free space. As an example, we look at a wavelength of 532 nm and compare all three MDPC. Fig. 4 shows the image formed by the three lenses and also the case of the transparent metal MDPC, but without evanescent components (diffraction limited). The slits (40 nm wide with a center to center spacing of 140 nm) are placed at the entrance of the lens (but in free space) and the image plane is located 50 nm beyond the end face of each lens. Although each lens contains six, 22 nm thick Ag films, the lenses have slightly different lengths due to the different geometries. The result is that the distance from the object plane to the image plane is 357 nm, 391 nm, and 392 nm for the symmetric, transparent metal, and periodic MDPC, respectively. The transparent metal MDPC provides >95% contrast for the two slits at a slit separation of $\lambda/4$. The other two lenses are only slightly better than the diffraction limited case.

It has been noted previously that the interference of propagating waves and evanescent waves can cause circulation in the Poynting vector[5]. Remnants of the vortices can be seen in Fig. 4 for the slightly negative components of $S_z$ at a distance of ~200 nm from the center of the slits. The presence of vortices is an indication that evanescent waves are contributing to the resolution of the lens.

Images similar to those shown in Fig. 4 were calculated throughout the transparency band of the transparent metal lens. It was found that over the wavelength range of 500 nm to 650 nm, the transparent metal lens could resolve two 40 nm wide slits with a contrast >80% and slit separation of <$\lambda$/2.5 where lambda is the free space wavelength for the incident radiation. At most wavelengths, the slit separation was ~$\lambda$/4. In addition, the transmittance for the normal incidence propagating waves was ~ 50% over the super-lensing band, Fig. 1.

It is worthwhile mentioning that the single Fabry-Perot cavity (Ag/GaP/Ag) has good transparency for the propagating and evanescent waves in the region of 575 nm to 625 nm. Although the single cavity lens is short, only 79 nm, it provides the basic building block of the transparent metal lens.

In summary, we have taken a different approach to the development of a super-resolving lens by using a transparent metal structure known to have a broadband transmittance for propagating waves and a low electrical sheet resistance due to the high metallic content. The combination of coupled cavities and anti-reflection coatings yield transparent structures containing 10 optical skin depths of metal. The transparent metal lens provides a substantial improvement in the transparency of propagating and evanescent waves compared with a periodic or symmetric structure. The improved transmittance occurs over a broad range of wavelengths and Fourier components. Based on the simulations in Ref. 11, super-resolving transparent metals containing 40 optical skin depths of Ag should be possible. Finally, the transparent metal approach is based on relatively thick metal films that can be fabricated by traditional deposition techniques.

Figure Captions

Fig. 1) Normal incidence transmittance of a single Fabry-Perot cavity, Ag/GaP/Ag (22 nm/35 nm/22 nm); Periodic MDPC [6 periods of Ag/Gap (22 nm/35 nm)]; a Symmetric MDPC [5.5 periods of Ag/GaP (22 nm/35 nm)]; and a Transparent Metal MDPC [5.5 periods of Ag/Gap (22 nm/35 nm) with 17 nm thick GaP AR coatings on the entrance and exit faces].

Fig. 2a) Transmittance of the Fourier components at a wavelength of 532 nm. 2b) Transmittance versus wavelength for a single Fourier component at $k_x/k_o=3$.

Fig. 3) Topographic 3-dimensional plot of the transmittance versus wavelength and $k_x/k_o$ for the transparent metal MDPC and the Periodic MDPC. The lowest contour line indicates a transmittance of 16%. Note that the contour lines are not uniformly spaced to include the wide range of transmittance values.

Fig 4) Image formed by the lenses for 2 slit sources at a wavelength of 532 nm. The center to center slit separation is λ/4. The image plane is 50 nm from the end of the lens. The diffraction limited case is for the transparent metal but without the evanescent components included in the calculation.

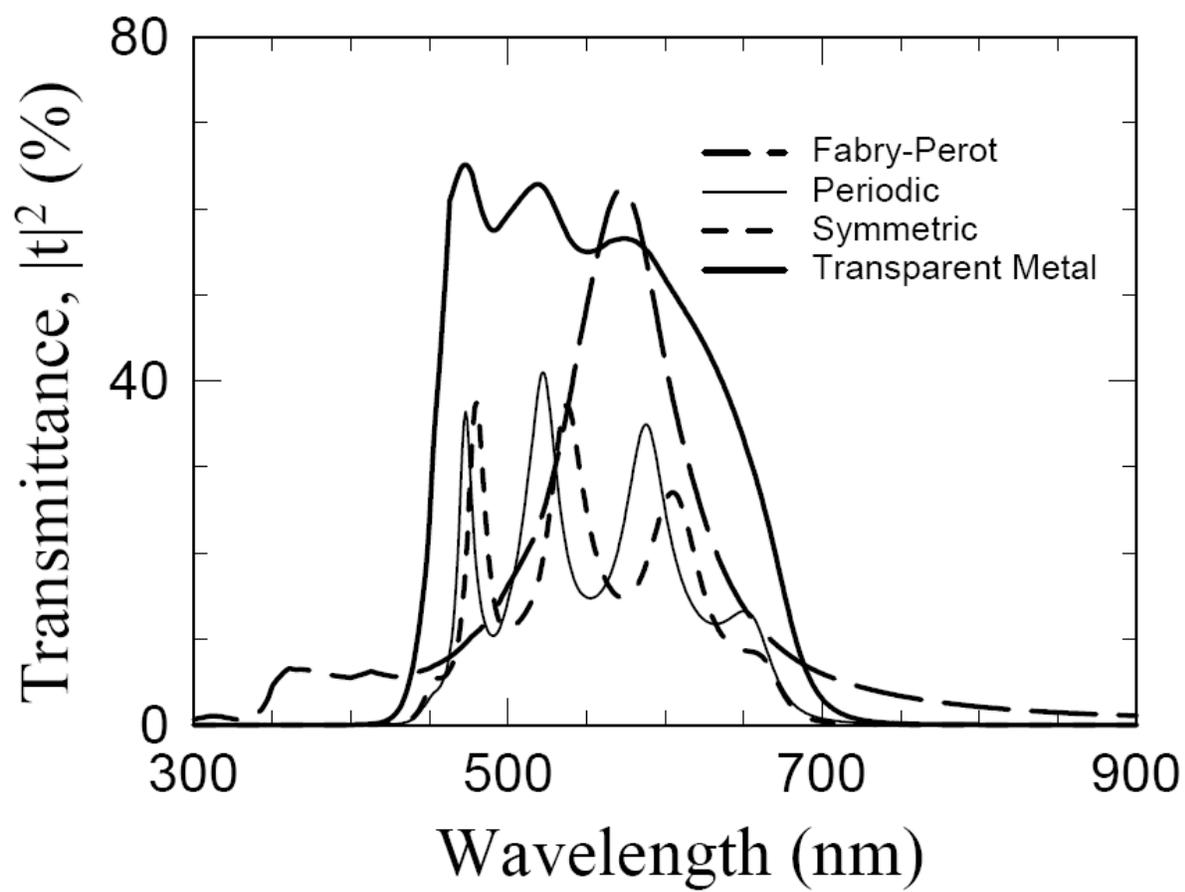

FIG. 1

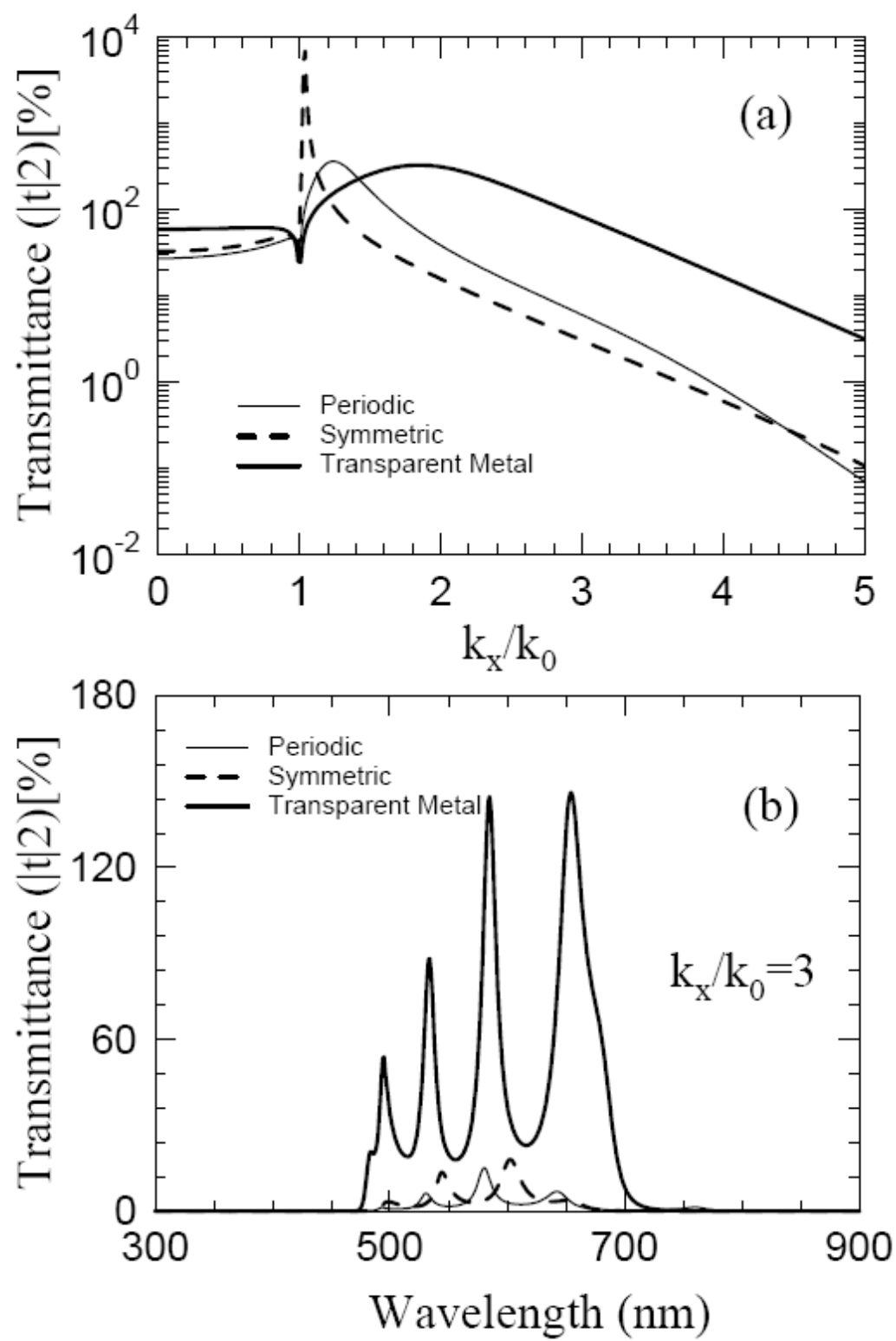

FIG. 2

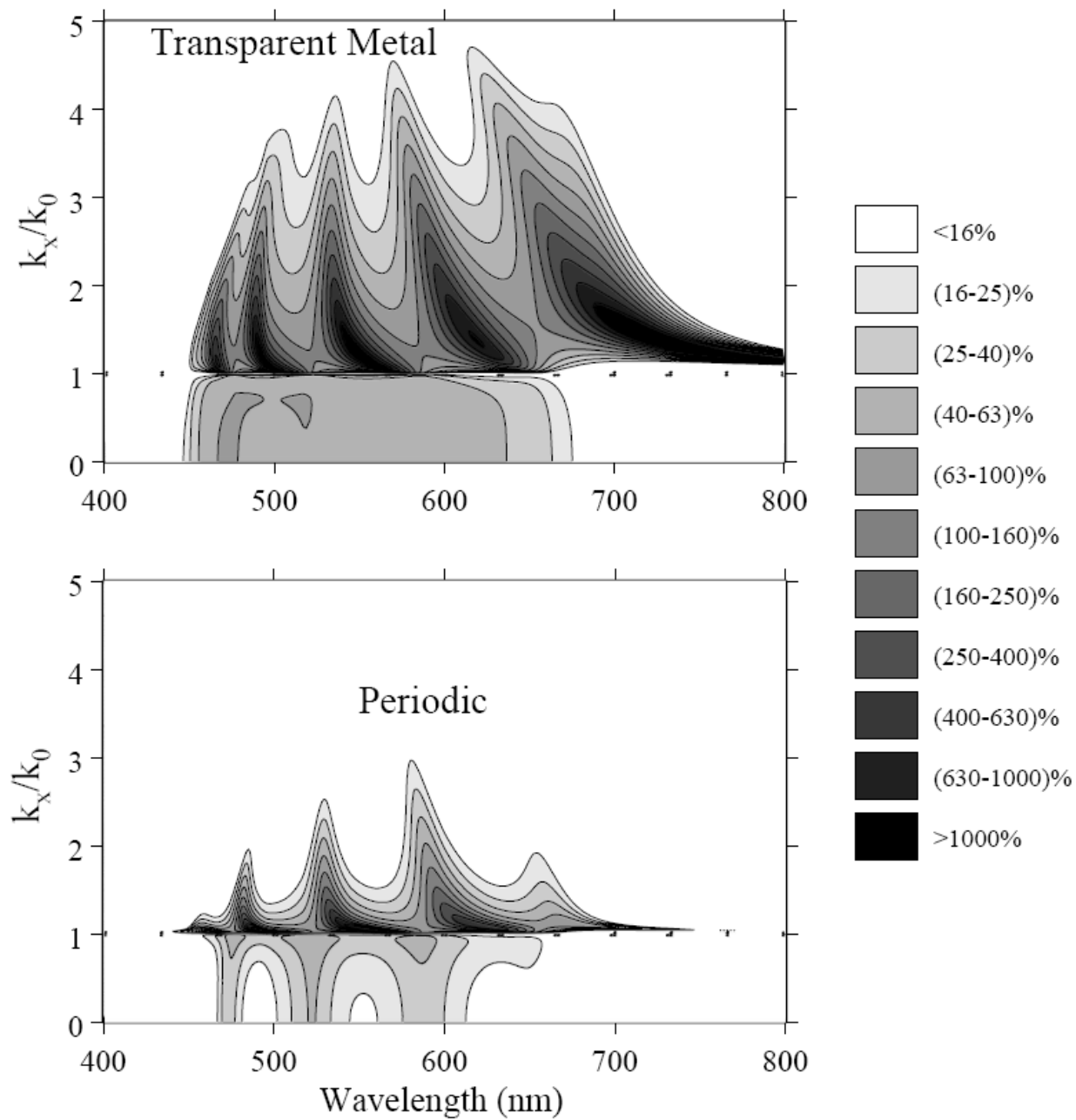

FIG. 3

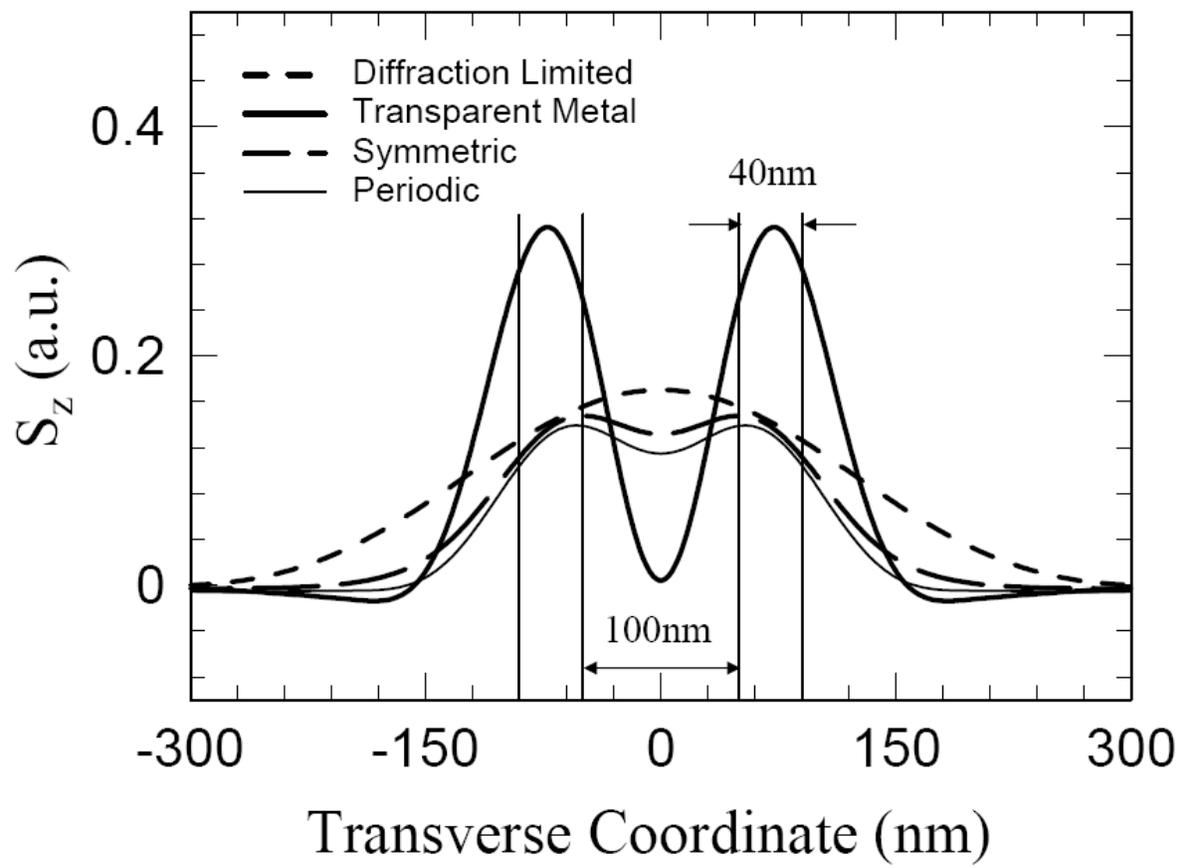

FIG. 4